\documentclass[11pt,letterpaper,oneside]{article}
\usepackage{graphicx}
\usepackage{amssymb,amsmath,amsfonts}
\usepackage{bm}
\usepackage[a4paper,left=1in,top=1in,right=1in,bottom=1in,ignoreheadfoot]{geometry}
\usepackage{algorithm}
\usepackage{setspace}

\providecommand{\cP}{\mathcal{P}}
\renewcommand{\Re}{\mathbb{R}}

\providecommand{\E}{\mathbb{E}}

\providecommand{\bx}{{\bf x}}
\providecommand{\by}{{\bf y}}

\providecommand{\bbeta}{{\bm \beta}}
\providecommand{\btheta}{{\bm \theta}}
\providecommand{\bX}{{\bf X}}

\providecommand{\btau}{{\bm \tau}}
\providecommand{\tby}{\tilde{\by}}

\providecommand{\T}{\mathbb{T}}
\providecommand{\pd}{{\bm 1}_{\cP}}
\usepackage{threeparttable}
\usepackage{pgf}
\usepackage{tikz}
\usepackage{subfig}
\usetikzlibrary{arrows,automata}

\providecommand{\eqdef}{\stackrel{\mbox{\upshape\tiny def}}{=}}

\begin{document}
\setlength{\parindent}{0pt}
\setlength{\parskip}{\baselineskip}

\author{Anthony Lee \footnote{Oxford-Man Institute and Department of Statistics, University of Oxford, UK}
\and Francois Caron \footnote{INRIA Bordeaux Sud-Ouest and Institut de Math\'ematiques de Bordeaux, University of Bordeaux, France}
\and Arnaud Doucet \footnote{Institute of Statistical Mathematics, Japan and University of British Columbia, Department of Statistics and Department of Computer Science, Canada.}
\and Chris Holmes \footnote{Department of Statistics and Oxford-Man Institute, University of Oxford, UK}
}
\title{A Hierarchical Bayesian Framework for Constructing Sparsity-inducing Priors}

\maketitle

\begin{abstract}
Variable selection techniques have become increasingly popular amongst statisticians due to an increased number of regression and classification applications involving high-dimensional data where we expect some predictors to be unimportant. In this context, Bayesian variable selection techniques involving Markov chain Monte Carlo exploration of the posterior distribution over models can be prohibitively computationally expensive and so there has been attention paid to quasi-Bayesian approaches such as maximum {\it a posteriori} (MAP) estimation using priors that induce sparsity in such estimates. We focus on this latter approach, expanding on the hierarchies proposed to date to provide a Bayesian interpretation and generalization of state-of-the-art penalized optimization approaches and providing simultaneously a natural way to include prior information about parameters within this framework. We give examples of how to use this hierarchy to compute MAP estimates for linear and logistic regression as well as sparse precision-matrix estimates in Gaussian graphical models. In addition, an adaptive group lasso method is derived using the framework.
\end{abstract}

\section{Introduction}
There has been recent interest in sparse estimates for coefficients in regression problems, with this problem often termed variable selection in the literature. To this end, a variety of approaches have been proposed in both the statistics and signal processing literatures. Most of the computationally tractable approaches are the solutions of penalized optimization problems associated with regularization of the coefficients in likelihood optimization. Although not truly Bayesian approaches, often they can be interpreted as maximum {\it a posteriori} (MAP) estimates associated with the posterior density of the coefficients where the prior induces the regularization used in the optimization routine.

Denoting the coefficients by $\bbeta \in \Re^p$, a popular family of these computing estimates as solutions to penalized optimization problems involving the log-likelihood of the data given $\bbeta$ and $\ell_q$ penalization on the coefficients with $0 \leq q \leq 1$. When $q$ is in this range, the solutions are sparse for large enough values of multiplicative penalization weights. When $q \geq 1$, the penalization is additionally convex, a property that has made the choice of $q = 1$ particularly suitable when the log-likelihood is concave as this leads to a unique global maxima for the objective function.

Beginning with \cite{lasso}, it has become popular practice to use $\ell_1$-regularization on each component of $\bbeta$. However, use of identical penalization on each coefficient, e.g. $\lambda\sum^p_{j=1}|\beta_j|$ can lead to unacceptable bias in the resulting estimates \cite{penlike}, which has motivated use of sparsity-inducing non-convex penalties despite the increased difficulty in computing the resulting estimates. In particular, this has led to the adoption of ``adaptive'' methods \cite{adaptive_lasso,one_step} in the statistics literature and iteratively reweighted methods \cite{rwl1,rwl2} in the signal processing literature. 

We propose a hierarchical prior for $\bbeta$ that amounts marginally to a sparsity-inducing, non-convex penalty in MAP estimation. Further, the specific hierarchy gives rise to an expectation-maximization (EM) algorithm \cite{em} that is essentially an iteratively reweighted $\ell_q$-minimization algorithm. In one case, the algorithm corresponds to the iteratively reweighted $\ell_1$-minimization algorithm and has been independently suggested in both \cite{garrigues} and \cite{cevher}. Our hierarchical formulation of the prior, in contrast, allows users to incorporate prior information about different coefficients and allows flexibility in grouping variables together. For example, the framework gives immediately an adaptive version of the group lasso algorithm proposed in \cite{group_lasso}.

\section{The hierarchical adaptive lasso (HAL)}
We are interested in prior distributions for $\bbeta$ in a general regression settings. Let $\T_k \eqdef \{1,\ldots,k\}$. We are given $n$ observations $\{y_i\}^n_{i=1}$ and associated with each observation a vector of covariates $\bx_i \in \Re^p$ for $i \in \T_n$. We assume that the conditional distribution of each $y_i$ is independent given $\bx_i$ and has density $f(y | \bx, \bbeta,\theta)$, where $\bbeta \in \Re^p$ and $\theta \in \Theta$ parametrize the distribution of $y$ conditional on $\bx$. Defining $\by \eqdef (y_1,\ldots,y_n)' \in \Re^n$ and $\bX \eqdef (\bx'_1,\ldots,\bx'_n)' \in \Re^{n \times p}$, the conditional distribution of all of the observations is then given by $f(\by | \bX, \bbeta, \theta) \eqdef \prod^n_{i=1}f(y_i|\bx_i,\bbeta,\theta)$. We are primarily interested in the parameter $\bbeta$ and assume that each component $\beta_j$ has special meaning when equal to 0.

While a Bayesian approach would usually suggest approximating the posterior density
$$p(\bbeta | \by, \bX, \theta) \propto f(\by | \bX, \bbeta, \theta) p(\bbeta|\theta),$$
we focus here on MAP (point) estimates of $\bbeta$ since these are computationally easier to compute, especially when $f(\by | \bX, \bbeta, \theta)$ and $p(\bbeta|\theta)$ are concave, and their use is not uncommon when $p$ and/or $n$ are large. MAP estimates are computed by solving the optimization problem
$$\hat{\bbeta}_{MAP} = \arg \max_{\bbeta} f(\by | \bX, \bbeta, \theta) p(\bbeta | \theta)$$
or, equivalently
$$\hat{\bbeta}_{MAP} = \arg \max_{\bbeta} \log f(\by | \bX, \bbeta, \theta) + \log p(\bbeta | \theta)$$
The $\log p(\bbeta | \theta)$ can be thought of as a penalization term in optimizing the log-likelihood of the data $\log f(\by | \bX, \bbeta, \theta)$.

\subsection{Generalized t-distribution prior}
\label{section:bpf}
We propose a hierarchical approach to constructing priors for $\bbeta$. At the lowest level, we give each element $\beta_j$ of $\bbeta$ an independent normal prior with mean 0 and variance $\sigma_j^2$, ie. $p(\bbeta|\sigma^2_{1:p}) = \prod^p_{j=1}p(\beta_j|\sigma_j^2)$ where $\beta_j | \sigma_j^2 \sim N(0,\sigma_j^2)$. If we leave $\sigma_j^2$ for $j \in \T_p$ as hyperparameters, computing the resulting MAP estimate corresponds to $\ell_2$-penalized optimization of the log-likelihood.

If, instead, we model each $\sigma_j^2$ as being drawn from an exponential distribution with mean $2\tau_j^2$ we obtain a double-exponential distribution for $\beta_j$ after $\sigma_j^2$ has been integrated out, ie.
$$p(\beta_j|\tau_j) = \frac{1}{2\tau_j}\exp(-\frac{|\beta_j|}{\tau_j})$$
Computing the MAP estimate associated with this prior corresponds to $\ell_1$-penalized optimization and the solution itself is identical to the LASSO estimate when $f$ is a multivariate Gaussian with mean $\bX\bbeta$. This prior has become popular in recent years for variable selection since it induces sparsity in $\hat{\bbeta}_{MAP}$ for small enough values of $\tau_j$.

We propose adding another level of hierarchy to the prior by having separate random variables $\tau_j$ for each $j \in \T_p$ and placing inverse-gamma priors on each $\tau_j$. Indeed, if we let $\tau_j \sim IG(a_j,b_j)$ we obtain
\begin{align}
\label{eqn:hal_prior}
p(\beta_j|a_j,b_j) = \frac{a_j}{2b_j}\left ( \frac{|\beta_j|}{b_j} + 1 \right )^{-(a_j+1)}
\end{align}
after integrating out $\tau_j$, which we call the hierarchical adaptive lasso (HAL) prior since one can compute MAP estimates using this prior with a type of adaptive lasso algorithm, as can be seen in Section \ref{section:compute_map}. This is the density of a generalized t-distribution. Computing the MAP estimate associated with this prior corresponds to logarithmic penalization of the log-likelihood. From a Bayesian modelling perspective, the introduction of a distribution over the $\tau_j$ is a natural way to resolve the issue of believing that there are significant differences in the sizes of the coefficients of $\bbeta$ that cannot be modelled as having come from a distribution with as thin tails as a Laplace distribution.

\subsection{Computing MAP estimates}
\label{section:compute_map}
The optimization problem associated with the generalized t-distribution prior is not concave. However, one can find local modes of the posterior using the EM algorithm with the $\btau = \tau_{1:p}$ as latent variables. Indeed, each iteration of EM takes the form
$$\bbeta^{(t+1)} = \arg \max_{\bbeta} \log f(\by | \bX, \bbeta, \theta) + \int \log [p(\bbeta|\btau)] p(\btau | \bbeta^{(t)}, a, b) d\btau$$

The conjugacy of the inverse-gamma distribution with respect to the Laplace distribution gives
$$\tau_j | \beta_j^{(t)}, a_j, b_j \sim IG(a_j+1,b_j + |\beta_j|)$$
and with $p(\bbeta|\btau) = \prod^p_{j=1} p(\beta_j|\tau_j) = \prod^p_{j=1} 1/(2\tau_j)\exp(-|\beta_j|/\tau_j)$ yields
$$\bbeta^{(t+1)} = \arg \max_{\bbeta} \log f(\by | \bX, \bbeta, \theta) - \sum^p_{j=1} |\beta_j| \int \frac{1}{\tau_j} p(\tau_j | \beta_j^{(t)},a_j,b_j) d\tau_j$$
where the expectation of $1/\tau_j$ given $\tau_j \sim IG(a_j+1,b_j+|\beta^{(t)}_j|)$ is $(a_j+1)/(b_j + |\beta^{(t)}_j|)$.

As such, one can find a local mode of the posterior $p(\bbeta | \by, \bX, \bbeta, \theta)$ by starting at some point $\bbeta^{(0)}$ and then iteratively solving
$$\bbeta^{(t+1)} = \arg \max_{\bbeta} \log f(\by | \bX, \bbeta, \theta) - \sum^p_{j=1} w_j^{(t)}|\beta_j|$$
where
$$w_j^{(t)} = \frac{a_j+1}{b_j + |\beta^{(t)}_j|}$$

It is clear that for large enough values of $a_j$ and small enough values of $b_j$ that the MAP estimates obtained by the EM algorithm are sparse. In fact, any posterior mode with this prior corresponds to a weighted lasso solution, which is sparse when the penalization through $\{(a_j,b_j)\}^p_{j=1}$ is large enough.

\subsubsection{Oracle properties}
In the penalized optimization literature, some methods are justified at least partially by their possession of the oracle property: that for appropriate parameter choices, the method performs just as well as an oracle procedure in terms of selecting the correct variables and estimating the nonzero coefficients asymptotically in $n$. Using the HAL prior in Theorem 5 of \cite{one_step} gives us the oracle property if $a_j \rightarrow \infty$ and $n^{-1/2}a_j \rightarrow 0$ as $n \rightarrow \infty$. It is worth remarking that this property requires our prior on $\bbeta$ to depend on the number of observations, which is atypical in Bayesian inference. Intuitively, $a_j$ needs to increase as $n$ increases to ensure that the solution remains sparse whilst it cannot increase too quickly or consistency is lost. As pointed out in \cite{penlike}, this trade off is impossible to accomplish with the LASSO.

\subsection{Generalizations and extensions}
\subsubsection{Exponential power family}
\label{section:EP}
One can model $\beta_j$ more generally as coming from an exponential power distribution instead of a Laplace distribution. In this case, we can write
$$p(\beta_j|\eta_j,q) = \frac{1}{2\eta_j^{1/q}\Gamma(1+1/q)}\exp\left ( -\frac{|\beta_j|^q}{\eta_j}\right )$$
With an inverse-gamma prior on $\eta_j$, which enjoys conjugacy with respect to the exponential power distribution, we obtain
$$p(\beta_j|a_j,b_j,q) = \frac{\Gamma(a_j+1/q)}{2\Gamma(a_j)\Gamma(1+1/q)b_j^{1/q}}\left ( \frac{|\beta_j|^q}{b_j} +  1\right )^{-(a_j+1/q)}$$
Use of this prior results in the same algorithm but with the weights given by
$$w_j^{(t)} = \frac{a_j+1/q}{b_j + |\beta^{(t)}_j|^q}$$

The use of an exponential power prior can be motivated hierarchically as a scale mixture of normal distributions for $q \in [1,2)$ \cite{smn} or as a scale mixture of uniform distributions for $q \in (1,\infty)$ \cite{smu}. For $q \in (0,1)$ this distribution is still defined but it does not have the same interpretation as when $q \geq 1$ and additionally has a non-concave density which complicates computation of posterior modes. The choice $q=2$ corresponds to a normal distribution and after marginalizing out $\eta$ it gives a scaled $t$-distribution with $2a_j$ degrees of freedom and scale $\sqrt{b_j/a_j}$. This choice leads to a hierarchical adaptive $\ell_2$-regularized method that may be suitable for problems in which prediction instead of variable selection is more important.

\begin{figure}[pth]
\center
\subfloat[LASSO]{
\includegraphics[scale=0.2]{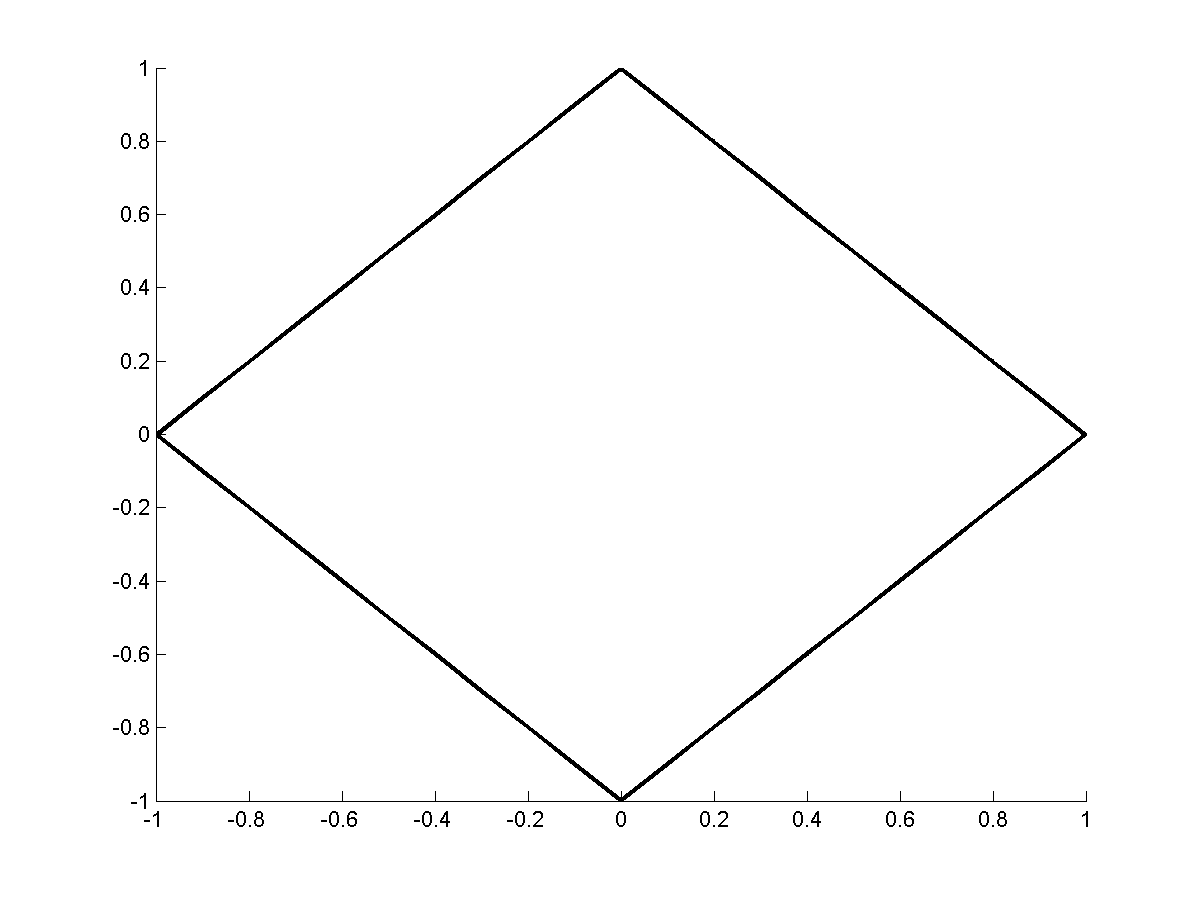}
}
\subfloat[HAL]{
\includegraphics[scale=0.2]{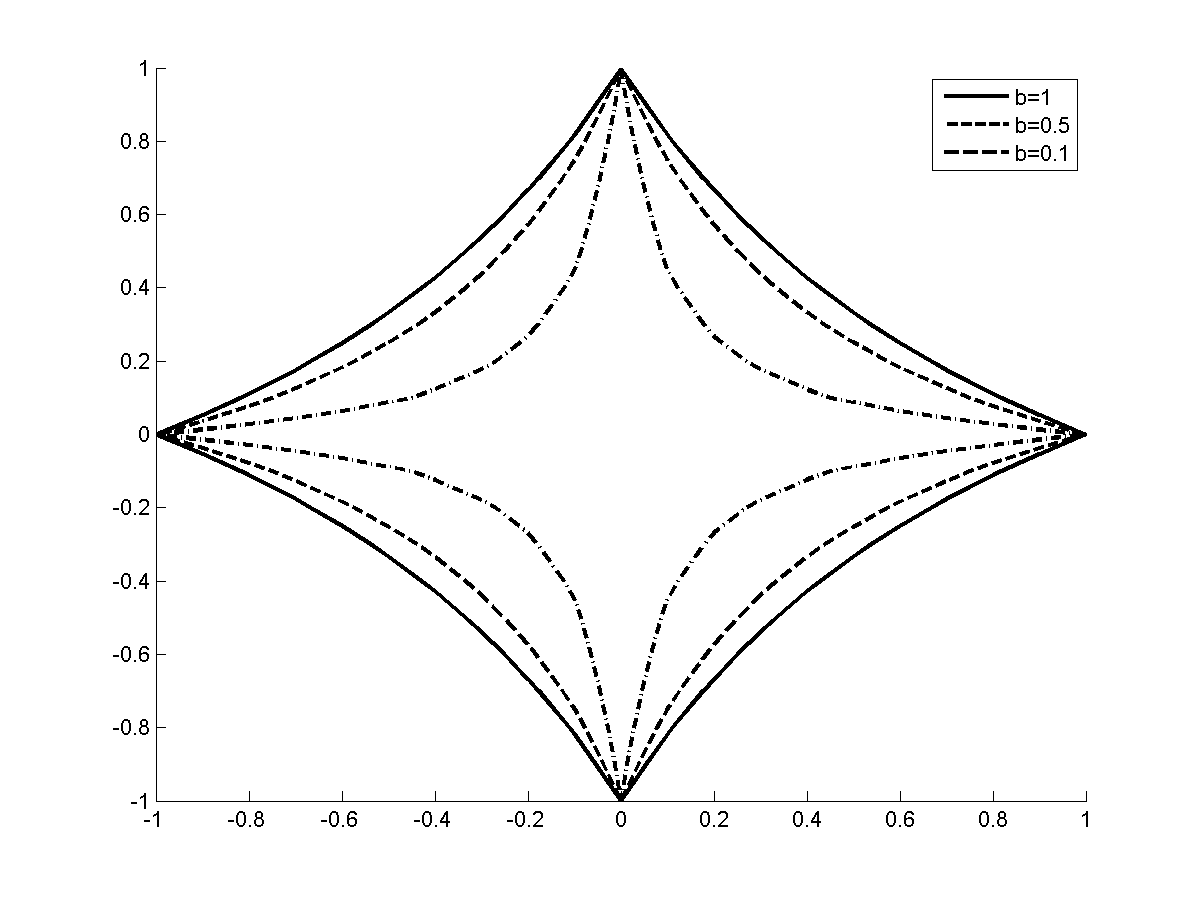}
}
\subfloat[HAR]{
\includegraphics[scale=0.2]{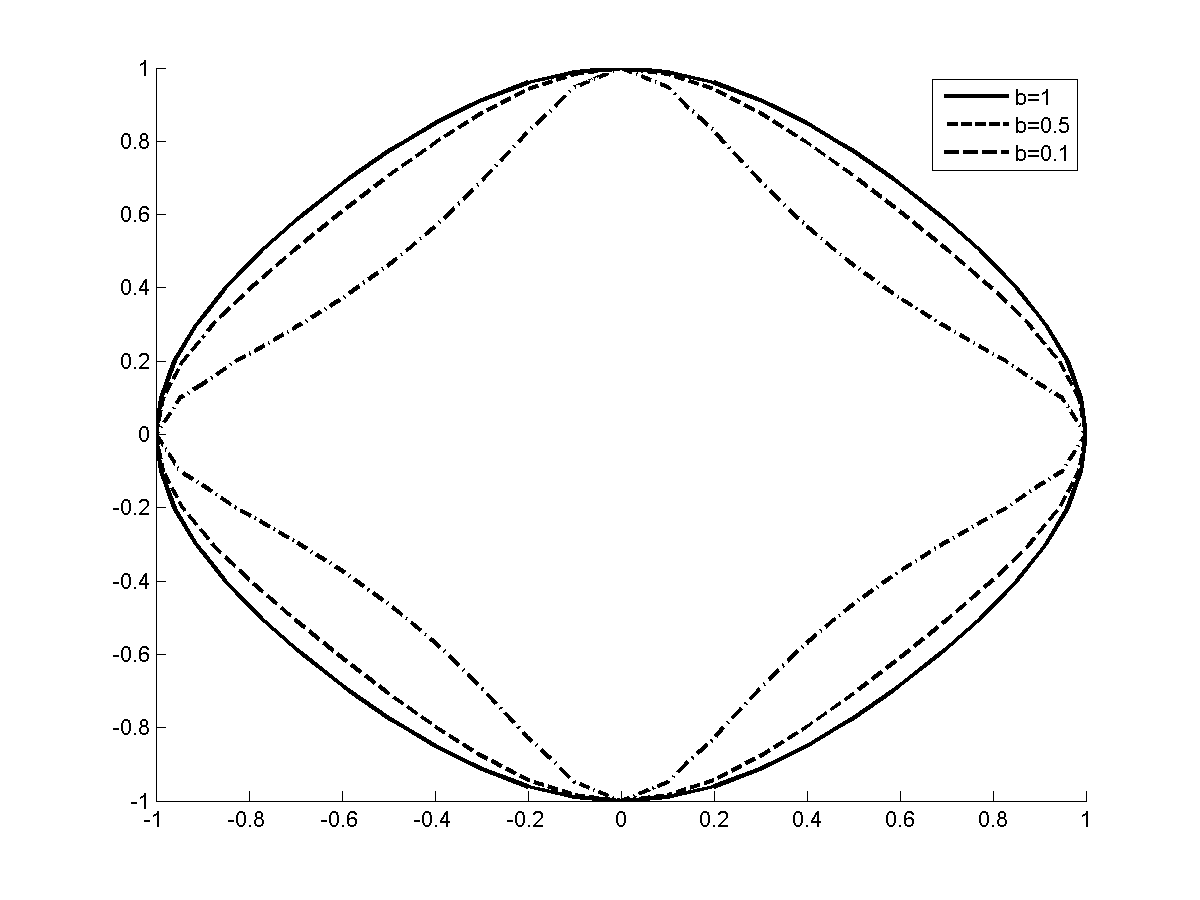}
}
\caption{Two-dimensional contour plots of the penalties, ie. the negative log density, associated with the priors.}
\label{fig:2d_contour}
\end{figure}

\begin{figure}[pth]
\center
\subfloat[LASSO]{
\includegraphics[scale=0.2]{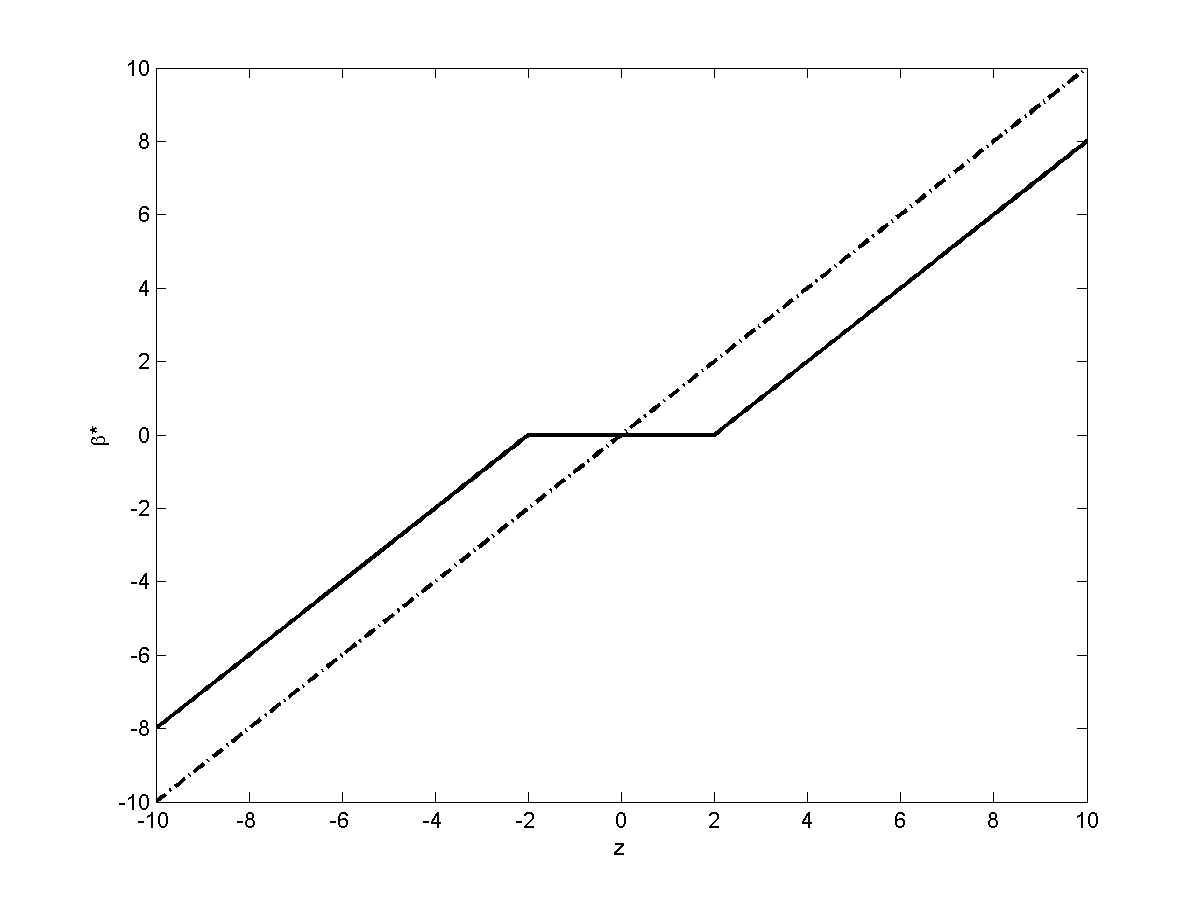}
}
\subfloat[HAL]{
\includegraphics[scale=0.2]{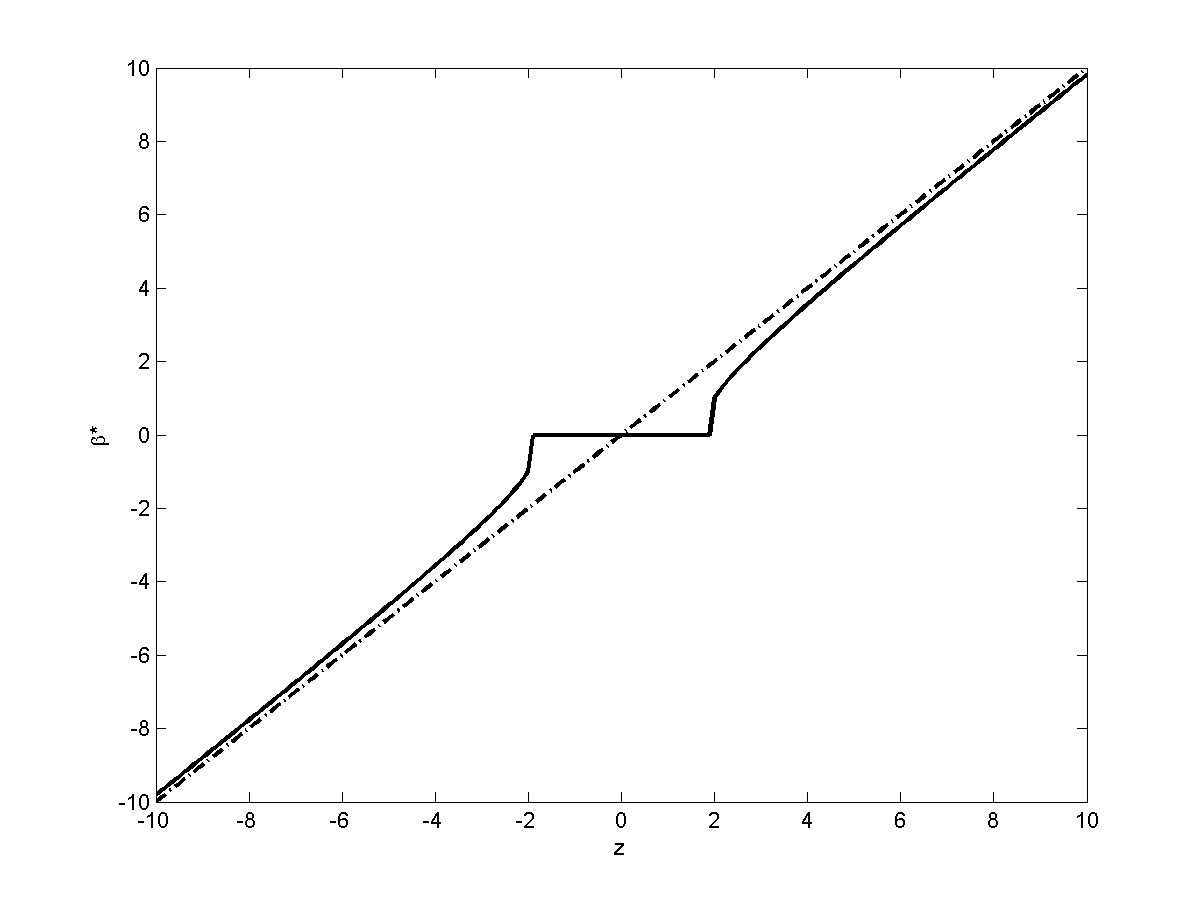}
}
\subfloat[HAR]{
\includegraphics[scale=0.2]{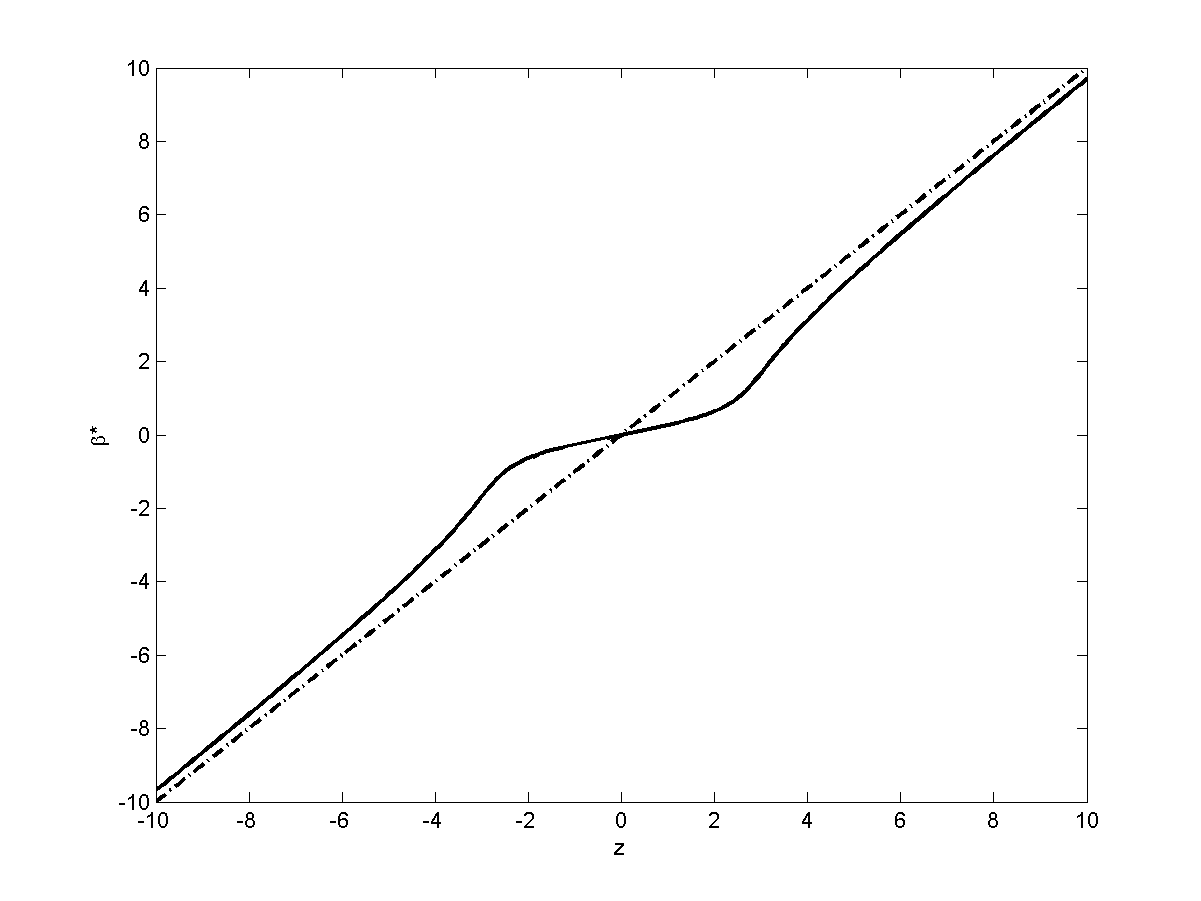}
}
\caption{Threshold plots associated with the priors.}
\label{fig:thresholds}
\end{figure}

Contour plots of the negative log density of the joint prior for two variables are given in Figure \ref{fig:2d_contour} and thresholding plots associated with the priors are given in Figure \ref{fig:thresholds}. The contour plots show graphically how the LASSO and HAL approaches give sparse solutions whilst the hierarchical adaptive ridge (HAR) prior, corresponding to $q = 2$, gives non-sparse solutions. The thresholding plots show that whilst the LASSO significantly biases even large coefficients, the HAL and HAR do not.

\subsubsection{The hierarchical adaptive group lasso}
\label{section:GL}
The hierarchical framework allows us to group variables together by making them dependent on a shared variable higher up in the hierarchy. For example, letting $g : \T_p \rightarrow \T_{K}$ be a function mapping variables to one of $K$ groups and $n_i$ be the number of variables in group $i$ we can use the following model:
\begin{align*}
\beta_j | \sigma^2_{g(j)} &\sim N(0,\sigma^2_{g(j)}), j \in \T_p \\
\sigma^2_{i} | \tau_{i} &\sim G \left ( \frac{n_i+1}{2},2\tau_{i}^2 \right ), i \in \T_{K} \\
\tau_{i} | a_{i},b_{i} &\sim IG(a_{i},b_{i}), i \in \T_{K} \\
\end{align*}

With $G_i = \{j : g(j) = i\}$, this gives
$$p(\beta_{G_i}|\tau_i) = \frac{(2\tau_i)^{-n_i}\pi^{-(n_i-1)/2}}{\Gamma((n_i+1)/2)}\exp(-\frac{\sqrt{\sum_{j \in G_i}{|\beta_j|^2}}}{\tau_i})$$
and so $\tau_i | \beta_{G_i},a_i,b_i \sim IG(a_i + n_i, b_i + \sqrt{\sum_{j \in G_i}{|\beta_j|^2}})$. The corresponding iterative procedure is then
$$\bbeta^{(t+1)} = \arg \max_{\bbeta} \log f(\by | \bX, \bbeta, \theta) - \sum^{K}_{i=1} w_{i}^{(t+1)}||\beta_{G_i}||_2$$
where
$$w_{i}^{(t+1)} = \frac{a_i + n_i}{||\beta^{(t)}_{G_j}||_2 + b_i}$$

The marginal prior on $\beta_{G_i}$ has the density
$$p(\beta_{G_i}|a_i,b_i) = \frac{(2b_i)^{-n_i}\pi^{-(n_i-1)/2}\Gamma(n_i+a_i)}{\Gamma((n_i+1)/2)\Gamma(a_i)} \left ( \frac{||\beta^{(t)}_{G_i}||_2}{b_i} + 1 \right )^{(-a_i-n_i)}$$
but this density is never evaluated in the EM algorithm.

A related problem to grouped variable selection is known as multi-task learning within the machine learning literature, where one wants to solve for $\btheta \eqdef \{\bbeta^{(i)}\}^L_{i=1}$ in a variety of $L$ related regression models. One approach is to solve the optimization problem
$$\hat{\btheta}_{MAP} = \arg \max_{\btheta} \sum^L_{i=1} \log f_i(\by_i | \bX_i, \bbeta^{(i)}) + \sum^p_{j=1} \lambda_j ||\bbeta_j||_2$$
where $\bbeta_j \eqdef (\bbeta_j^{(1)},\ldots,\bbeta_j^{(L)}) \in \Re^L$ \cite{mtfs}. This type of regularization can be derived using the same hierarchical prior used in the group lasso where the coefficients relating to the same covariate are `grouped' together to promote sparsity across the individual $\bbeta$ estimates, ie. a covariate is selected in all the related models or in none of the models. As such, an adaptive version of this multi-task learning approach follows the same form as the hierarchical adaptive group lasso.

\subsubsection{Matrix priors}
For the purpose of covariance matrix estimation, $\ell_1$-regularization has been used on entries of the precision matrix $\Omega$ of a Gaussian graphical model \cite{ggm, graphical_lasso}. This corresponds to MAP estimation using Laplace priors on each $\Omega_{ij}$ for $i \leq j$. We can incorporate this type of prior within our framework by placing inverse Gamma priors on the scale parameters of each Laplace distribution. We have
$$p(\Omega_{ij} | \tau_{ij}) = \frac{1}{2\tau_{ij}} \exp(-\frac{|\Omega_{ij}|}{\tau_{ij}})$$
with $p(\Omega | \tau) = \prod^p_{i=1} \prod^p_{j=i} p(\Omega_{ij} | \tau_{ij})$ and $\tau_{ij} \sim IG(a_{ij},b_{ij})$. Note that in this formulation the prior on $\Omega$ is non-zero for non-positive-definite values. This allows us to specify
$$p(\btau | \Omega, A, B) = \prod^p_{i=1} \prod^p_{j=i} IG(\tau_{ij} ; a_{ij} + 1, b_{ij} + |\Omega_{ij}|)$$

One can have the likelihood of observed data $Y$, $p(Y|\Omega)$ be zero if $\Omega$ is not symmetric positive-definite. In this case, the posterior and hence the MAP estimate are equivalent to the case where the prior takes the form
$$p(\Omega|A,B) = \frac{\pd(\Omega) p(\Omega | A, B)}{\int \pd(\Omega) p(\Omega | A, B) d\Omega}$$
since in both cases we have
$$p(\Omega|Y,A,B) = \frac{\pd(\Omega) p(Y|\Omega)p(\Omega|A,B)}{\int \pd(\Omega) p(Y|\Omega) p(\Omega | A,B) d\Omega}$$
where $\cP$ is the set of symmetric positive-definite matrices. Note, however, that the positive-definite prior cannot be used to derive the EM algorithm central to our methodology since $\btau | \Omega, A,B$ is no longer a product of inverse-gamma distributions.

\subsubsection{The hierarchical lasso}
In some cases, one might be interested in having $\eta_j = \eta$ for all $j \in \T_p$ with $\eta \sim IG(a,b)$. In this case, one obtains a prior on $\bbeta$ of the form
\begin{align}
\label{eqn:hl_prior}
p(\bbeta | a, b, q) = \frac{\Gamma(a+p/q)}{2^p\Gamma(a)\Gamma(1 + 1/q)^p b^{p/q}}\left ( \frac{\sum^p_{j=1}|\beta_j|^q}{b} + 1 \right )^{-a-p/q}
\end{align}
which leads to the iterative procedure
$$\bbeta^{(t+1)} = \arg \max_{\bbeta} \log f(\by | \bX, \bbeta, \theta) - w^{(t+1)} \sum^{p}_{j=1} |\beta_{j}|^q$$
where
$$w^{(t+1)} = \frac{a + p/q}{b + \sum^{p}_{j=1} |\beta^{(t)}_{j}|^q}$$

In fact, a more general prior can be constructed by considering groupings of the coefficients such that $\beta_j \sim EP(\eta_i, q)$ for all $j \in G_i$, where $G_i$ is again the set of indices of coefficients in group $i$. A prior constructed in this fashion leads to the iterative procedure
$$\bbeta^{(t+1)} = \arg \max_{\bbeta} \log f(\by | \bX, \bbeta, \theta) - \sum^{K}_{i=1} w^{(t+1)}_i \sum_{j \in G_i}|\beta_{j}|^q$$
where
$$w_i^{(t+1)} = \frac{a_i + n_i/q}{b_i + \sum_{j \in G_i} |\beta^{(t)}_{j}|^q}$$

\subsubsection{Modifying the hierarchy}
The above examples are only a subset of the possible modifications to the hierarchy that are possible. Indeed, one of the benefits of a hierarchical approach is that one can flexibly group variables via the sharing of random variables. Graphical models for the exponential family generalization and the grouped variable generalization are given in Figure \ref{fig:graph} along with a discussion of their relationships to existing methods in Section \ref{section:gm}.

\subsection{Tuning the hyperparameters}
Use of the proposed framework relies on appropriate settings of the hyperparameters. For distributions of non-negative $Z$ with density
$$p(Z|\nu,b) = \frac{\nu - 1}{b} \left ( \frac{Z}{b} + 1 \right )^{-\nu}$$
the moments of $Z$ are given by
$$\E_p[Z^t] = \frac{b^t\Gamma(\nu - 1 - t)\Gamma(t+1)}{\Gamma(\nu -1)}$$
which allows one to pick hyperparameters that represent prior beliefs about the mean and variance of variables of interest, e.g. $|\beta_j|$ in the case of prior (\ref{eqn:hal_prior}) or $(\sum^p_{j=1}|\beta_j|^q)^{1/q}$ in the case of prior (\ref{eqn:hl_prior}).

Focusing on the hierarchical adaptive lasso prior, we note that in this case we have $\E[|\beta_j|] = b_j/(a_j-1)$, for $a_j > 1$. An observation on $a_j$ and $b_j$ is that when one increases both values but keeps $\E[|\beta_j|]$ constant, the tendency for the iterative scheme to set $\beta_j$ to zero is reduced since $w_j$ is upper-bounded by $(a_j+1)/b_j$. This observation could be used in a `tempered' optimization scheme as discussed in Section \ref{section:temper}, noting in particular that as $a_j \rightarrow \infty$, the prior approaches a Laplace distribution and so the posterior approaches unimodality.

\subsection{Issues with MAP estimation}
\label{section:temper}
There are many criticisms of MAP estimates in a Bayesian framework. We motivate use of such estimates for primarily computational reasons, since Bayesian variable selection methods tend to be prohibitively expensive when dealing with large data sets. Beyond the obvious problem of summarizing the posterior distribution over models with a point estimate, one problem is that MAP estimates are not Bayes estimators but instead a limit of Bayes estimators under the 0-1 loss function. While important, this issue is not addressed here. A perhaps more fundamental issue is that MAP estimates are not invariant under reparametrization. This issue can be rectified by finding the point that maximizes posterior density with the Jeffreys measure as the dominating measure \cite{jermyn,druilhet}, eg. for a likelihood $f(x|\theta)$ and prior $p(\theta)$, the parametrization-invariant MAP is given by
$$\theta_{MAP} = \arg \max_\theta f(x|\theta) p(\theta) |I(\theta)|^{-1/2}$$
where $I(\theta)$ is the Fisher information associated with $f(x|\theta)$.

An important issue with the MAP estimates obtained from our methodology is that the posterior is multimodal and there is no guarantee that one will obtain the global mode of the posterior as opposed to a local one. However, this is true of almost all non-convex penalized optimization approaches. In \cite{rwl2}, a suggestion is to start with high values of $b_j$ and reduce the values of $b_j$ once the algorithm has converged. In principle, this can be done with both $a_j$ and $b_j$, noting that such an algorithm will still find a local mode of the posterior and this `tempering' of the posterior during optimization can affect which mode is chosen. We do not investigate this further but note that characterization of the modes obtained using such a process is an interesting open question.

\section{Related approaches}
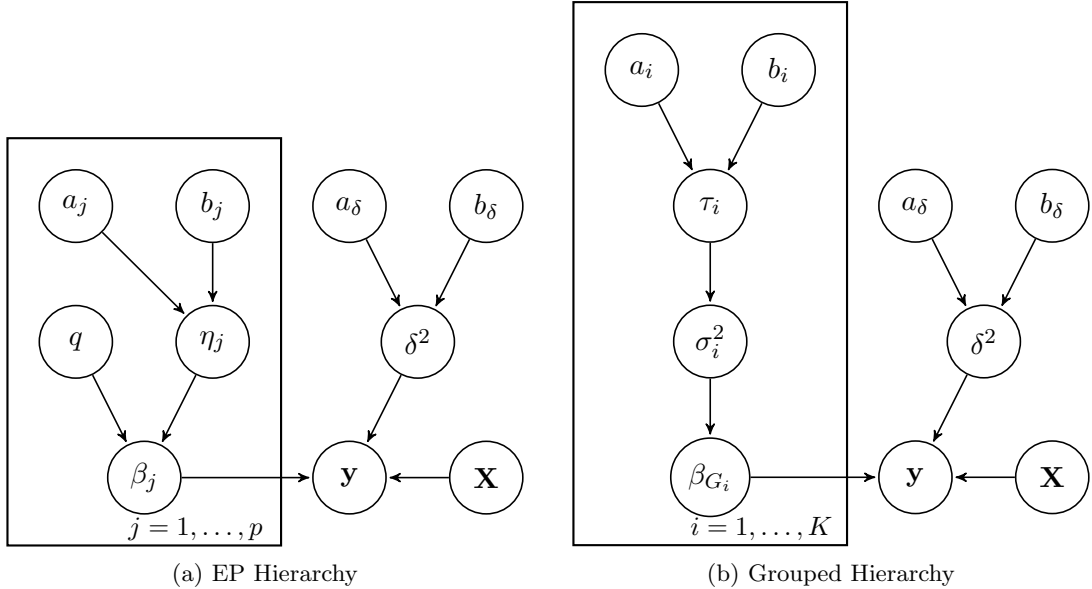
\begin{figure}
\center
\subfloat[EP Hierarchy]{
\begin{tikzpicture}[scale=.9,->,>=stealth',shorten >=1pt,auto,node distance=2.8cm,semithick]
	\tikzstyle{every state}=[fill=white,draw=black,text=black]
  \tikzstyle{ann} = [draw=none,fill=none]
  \draw[thick] (-1,-1) rectangle (3cm,5cm); 
  \draw (0,4) node[state]         (A)                    {$a_j$};
  \draw (2,4)  node[state]         (B)        {$b_j$};
  \draw (2,2) node[state]         (C)        {$\eta_j$};
  \draw (0,2) node[state]         (D)        {$q$};
  \draw (1,0) node[state]         (E)    {$\beta_{j}$};
  \draw[draw=none,fill=none] (1.75,-0.75) node[ann] {\small $j=1,\ldots,p$};
  \draw (4,0) node[state]         (F) {$\by$};
  \draw (6,0) node[state]         (G) {$\bX$};
  \draw (5,2) node[state]         (H) {$\delta^2$};
  \draw (4,4) node[state]         (I) {$a_\delta$};
  \draw (6,4) node[state]         (J) {$b_\delta$};

  \path (A) edge              node {} (C)
        (B) edge              node {} (C)
        (D) edge              node {} (E)
        (C) edge              node {} (E)
        (E) edge              node {} (F)
        (G) edge              node {} (F)
        (I) edge              node {} (H)
        (J) edge              node {} (H)
        (H) edge              node {} (F);
\end{tikzpicture}
}
\quad
\subfloat[Grouped Hierarchy]{
\begin{tikzpicture}[scale=.9,->,>=stealth',shorten >=1pt,auto,node distance=2.8cm,semithick]
  \tikzstyle{every state}=[fill=white,draw=black,text=black]
  \tikzstyle{ann} = [draw=none,fill=none]
  \draw[thick] (-1,-1) rectangle (3cm,7cm); 
  \draw (0,6) node[state]         (A)                    {$a_i$};
  \draw (2,6)  node[state]         (B)        {$b_i$};
  \draw (1,4) node[state]         (C)        {$\tau_i$};
  \draw (1,2) node[state]         (D)     {$\sigma_i^2$};
  \draw (1,0) node[state]         (E)    {$\beta_{G_i}$};
  \draw[draw=none,fill=none] (1.75,-0.75) node[ann] {\small $i=1,\ldots,K$};
  \draw (4,0) node[state]         (F) {$\by$};
  \draw (6,0) node[state]         (G) {$\bX$};
  \draw (5,2) node[state]         (H) {$\delta^2$};
  \draw (4,4) node[state]         (I) {$a_\delta$};
  \draw (6,4) node[state]         (J) {$b_\delta$};

  \path (A) edge              node {} (C)
        (B) edge              node {} (C)
        (C) edge              node {} (D)
        (D) edge              node {} (E)
        (E) edge              node {} (F)
        (G) edge              node {} (F)
        (I) edge              node {} (H)
        (J) edge              node {} (H)
        (H) edge              node {} (F);
\end{tikzpicture}
}
\caption{Graphical model representations of the exponential power and the grouped variable hierarchies}
\label{fig:graph}
\end{figure}

The proposed approach, either in the hierarchical model or in the estimation step, is closely related to many approaches that have been suggested in the literature. One contribution of this paper is therefore to provide a Bayesian interpretation of existing methods and a flexible framework with which we can incorporate different models.

\subsection{Laplacian scale mixture distributions and compressible priors}
It has come to our attention that the HAL prior has been proposed independently in both \cite{garrigues} and \cite{cevher}. In the former, one obtains the same procedure from a majorization-minimization algorithm and in the latter from an EM algorithm. However, our derivation makes explicit the flexibility of the hierarchy and generalizes this prior to exponential power families, situations with grouped variables and positive-definite matrices, making particularly clear strategies for choosing hyperparameters.

\subsection{Weakly informative priors and non-convex penalization}
With $\beta_j \sim N(0,\sigma_j^2)$ and $\sigma_j^2 \sim IG(a_j,b_j)$ one obtains marginally a t-distribution for $\beta_j | a_j,b_j$. This corresponds to the idea of using weakly informative priors as in \cite{Gelman2008}. For the case where $\beta_j \sim \text{Laplace}(0,\tau)$ and $\tau \sim IG(a_j,b_j)$, ie. the hierarchical adaptive lasso, we can similarly think of the generalized t-distribution prior on $\beta_j$ after marginalizing out $\tau$ as a weakly informative prior. In fact, one can think of all of the priors proposed using the hierarchical approach in this work as weakly informative.

\subsection{Adaptive methods}
Within the statistical literature, the closest approach is perhaps the adaptive lasso \cite{adaptive_lasso}, whose implementation corresponds to a single step of the exponential power family generalization of the HAL with $\bbeta^{(0)}$ a root-$n$ consistent estimator of $\beta$ and $b_j \rightarrow 0$. As such, the adaptive lasso estimator can be thought of as taking an initial estimate and returning an estimate with higher posterior density given a logarithmic prior. Our method, on the other hand, finds a local mode of the posterior.

Similarly, the benefit of a polynomial form for the prior density is related to the motivation for the smoothly clipped absolute deviation penalty \cite{penlike}. Indeed, the penalization induced by the HAL prior grows slowly so that large values of $\beta_j$ are not unnecessarily biased while remaining continuous and sparse. The methods used to find local linear and local quadratic approximations (LLA and LQA) algorithms of \cite{penlike, one_step} are also closely related, being iteratively reweighted optimization algorithms with a different penalization.

\subsection{Iteratively reweighted $\ell_q$-minimization}
The basic HAL algorithm is clearly similar to the reweighted-$\ell_1$ approach proposed in \cite{rwl1}, which is identical except that the weights have the form
$$w_j^{(t)} = \frac{\lambda}{\epsilon + |\beta^{(t)}_j|}$$
which corresponds to a limiting case where $a_j \rightarrow \lambda - 1$ and $b_j$ is set to be small.

Similarly, the exponential-family generalization of the HAL algorithm is related to the family of approaches suggested in \cite{rwl2} for the various $\ell_q$-penalization norms. As such, the hierarchical model for $\bbeta$ gives an interpretation to the methods in the family of iteratively reweighted optimization solutions and, in particular, to the selection of additional  parameters $\epsilon$ and $\lambda$.

\subsection{Normal-Exponential-Gamma priors}
Our hierarchical prior differs from that suggested in \cite{griffin07} in that an inverse gamma prior is placed on $\tau_j^2$ as opposed to $\tau_j$. This difference in their work results in a posterior for $\bbeta$ for which it is difficult to obtain MAP estimates \cite{griffin10}, although this problem can be alleviated by novel fast methods for computation of the parabolic cylinder function \cite{murphy}. The marginal prior is a member of the generalized hyperbolic family. This difference also appears in \cite{bayesian_lasso}, although in that work the full posterior is explored using MCMC.

\subsection{A note on improper priors}
Consider the exponential power density
$$p(\beta_j|\eta_j,q) = \frac{1}{2\eta_j^{1/q}\Gamma(1+1/q)}\exp\left ( -\frac{|\beta_j|^q}{\eta_j}\right )$$
with the scale-invariant prior on $\eta_j$, $p(\eta_j) \propto 1/\eta_j$. The prior on $\beta_j$ after marginalizing out $\eta_j$ is then, regardless of $q$, improper with the form $p(\beta_j | q) \propto 1/|\beta_j|$. Since this is the same prior for $q = 1$, which we know will produce sparse $\bbeta$ and for $q = 2$, which is the prior proposed in \cite{figuerido}, this explains why the prior in \cite{figuerido} produces sparse results. However, it is worth noting that the posterior for $\bbeta$ using this prior is improper with unbounded density at $\bbeta = {\bm 0}$.

\subsection{Graphical Model}
\label{section:gm}
Figure \ref{fig:graph} gives graphical models for the hierarchies corresponding to the exponential power (EP) generalization of section \ref{section:EP} and the adaptive group lasso of section \ref{section:GL}. These models allow us to visualize the flexibility of the framework and the connections with related approaches. Indeed, for $q = 1$ one obtains the hierarchical adaptive lasso or, by setting $\eta_j$ to be a fixed hyperparameter, the standard lasso. Similarly, for $q=2$ one obtains hierarchical adaptive ridge regression or standard ridge regression. For the hierarchy with grouped variables, the similarity to the hierarchical adaptive lasso hierarchy is clear, suggesting that application-specific hierarchies could be developed that lead to iteratively reweighted methods.

\section{Examples}

\subsection{Linear regression}
In linear regression, the likelihood of $\by$ given $X$ and $\bbeta$ is given by
$$p(\by|\bX,\bbeta,\mu,\delta^2) = \frac{1}{(2\pi\delta^2)^{n/2}}\exp \left \{-\frac{1}{2\delta^2}(\tby_\mu - \bX\bbeta)^T(\tby_\mu - \bX\bbeta) \right \}$$
where $\tby_\mu \eqdef \by - \mu {\bf 1}_n$. If $X$ is standardized, we have $1_n^T X = 0$ and so we can put an improper prior on $\mu$ with $p(\mu) \propto 1$ and integrate it out so that
\begin{align}
p(\by|\bX,\bbeta,\delta^2) = \frac{1}{(2\pi\delta^2)^\frac{n-1}{2}\sqrt{n}} \exp \left \{ -\frac{1}{2\delta^2}(\tby - \bX \bbeta)^T(\tby - \bX \bbeta) \right \}
\end{align}
where $\bar{y} = \frac{1}{n}\sum^n_{i=1}y_i$ and $\tby = \by - \bar{y} 1_n$.

\subsubsection{Fixed $\delta^2$}
If $\delta^2$ is fixed, we proceed as expected. Note that in this case, the Jeffreys prior for $\bbeta$ is a uniform improper prior so no adjustment needs to be made to make the MAP estimate invariant.

To test the method, we simulated data using $\bbeta = (3,1.5,0,0,2,0,0,0)^T$, $\delta^2 = 1$ and $X \sim N(0,\Sigma)$ with $\Sigma_{i,j} = 0.5^{|i-j|}$. We then ran 1000 repetitions of the hierarchical adaptive lasso and the standard lasso on this problem with various settings of $(a,b)$ and $\tau$ respectively with the results given in Tables \ref{tab:lrd1_lasso}-\ref{tab:lrd1_hal}.ons of the hierarchical adaptive lasso and the standard lasso on this problem with various settings of $(a,b)$ and $\tau$ respectively with the results given in Tables \ref{tab:lrd1_lasso}-\ref{tab:lrd1_hal}.

\begin{table}[htp]
\small
\caption{Results for the LASSO (linear regression, $\delta = 1$) \vspace{10pt}}
\label{tab:lrd1_lasso}
\centering
\begin{tabular}{| c | c | c | c | c | c |}
\hline
$n$         & $\tau$        & avg. error      & \% correct   & avg. false positives & avg. false negatives \\
\hline
40          & 0.2           & 0.4137          & 9.7          & 1.808          & 0.0     \\
\hline
40          & 0.1           & 0.4817          & 36.7         & 0.89           & 0.0   \\
\hline
40          & 0.02          & 1.6732          & 90.0           & 0.089          & 0.015  \\
\hline
80          & 0.2           & 0.2872          & 2.8          & 2.519          & 0.0   \\
\hline
80          & 0.1           & 0.2931          & 20.0           & 1.3510         & 0.0   \\
\hline
80          & 0.02          & 0.8169          & 92.7         & 0.079          & 0.0      \\
\hline
\end{tabular}
\end{table}

\begin{table}[htp]
\small
\caption{Results for the HAL (linear regression, $\delta = 1$) \vspace{10pt}}
\label{tab:lrd1_hal}
\centering
\begin{tabular}{| c | c | c | c | c | c |}
\hline
$n$         & $(a,b)$       & avg. error      & \% correct   & avg. false positives & avg. false negatives \\
\hline
40          & $(1,0.1)$     & 0.3118          & 89.9         & 0.105          & 0.0     \\
\hline
40          & $(2,0.1)$     & 0.3044          & 98.1         & 0.019          & 0.0   \\
\hline
40          & $(2,0.05)$    & 0.3026          & 99.6         & 0.004          & 0.0   \\
\hline
80          & $(1,0.1)$     & 0.2191          & 81.3         & 0.079          & 0.0   \\
\hline
80          & $(2,0.1)$     & 0.2061          & 96.6         & 0.034          & 0.0   \\
\hline
80          & $(2,0.05)$    & 0.2038          & 98.8         & 0.012          & 0.0      \\
\hline
\end{tabular}
\end{table}

Both methods are capable of giving good results in this setting, which has a high signal-to-noise ratio. However, the reduction in average error is evident for the HAL, owing mainly to less penalization of the selected coefficients. We ran the same experiment but with $\delta = 3$, to test the algorithm with a lower signal-to-noise ratio with the results given in Tables \ref{tab:lrd3_lasso}-\ref{tab:lrd3_hal}. Again, the results for the HAL are typically superior to that for the LASSO. However, incorporating prior information leading to less penalization of $\beta_2$ and $\beta_5$ improves performance drastically. This type of prior information is likely to be necessary when we wish to include variables whose true coefficients are small.

\begin{table}[htp]
\small
\caption{Results for the LASSO (linear regression, $\delta = 3$) \vspace{10pt}}
\label{tab:lrd3_lasso}
\centering
\begin{threeparttable}
\begin{tabular}{| c | c | c | c | c | c |}
\hline
$n$         & $\tau$        & avg. error      & \% correct   & avg. false positives & avg. false negatives \\
\hline
40          & 1/6           & 1.9747          & 53.5          & 0.378          & 0.255     \\
\hline
40          & 0.125           & 2.3952         & 43.0          & 0.207           & 0.580   \\
\hline
40          & 0.125\tnote{$*$}  & 2.2117          & 93.9          & 0.005          & 0.057  \\
\hline
\end{tabular}
\begin{tablenotes}
 \item[$*$] $(\tau_2,\tau_5) = (0.25,0.25)$
\end{tablenotes}
\end{threeparttable}
\end{table}

\begin{table}[htp]
\small
\centering
\caption{Results for the HAL (linear regression, $\delta = 3$) \vspace{10pt}}
\label{tab:lrd3_hal}
\begin{threeparttable}
\begin{tabular}{| c | c | c | c | c | c |}
\hline
$n$         & $(a,b)$       & avg. error      & \% correct   & avg. false positives & avg. false negatives \\
\hline
40          & $(2,0.75)$     & 1.3407          & 56.2         & 0.274          & 0.352     \\
\hline
40          & $(2,0.1)$      & 1.7198          & 28.0         & 0.064          & 0.831   \\
\hline
40          & $(2,0.1)$\tnote{$*$} & 1.0224          & 95.9          & 0.004          & 0.038   \\
\hline
\end{tabular}
\begin{tablenotes}
 \item[$*$] $(a_2,b_2,a_5,b_5) = (2,2,2,2)$
\end{tablenotes}
\end{threeparttable}
\end{table}

\subsubsection{$\delta^2 \sim IG(a_\delta,b_\delta)$}
If we model $\delta^2 \sim IG(a_\delta,b_\delta)$, we can find that MAP estimate associated with the posterior density $p(\bbeta | \by,X)$, ie. with $\delta^2$ integrated out. To do so, we additionally include $\delta^2$ as a latent variable in the EM algorithm, noting that conditional on $\bbeta$, $\delta^2$ and $\btau$ are independent. Furthermore, we have $\delta^2 | \bbeta,X,\by \sim IG(a_\delta + (n-1)/2,b_\delta + 1/2(\tby - \bX \bbeta)^T(\tby - \bX \bbeta))$ For the hierarchical adaptive lasso, we iteratively solve
$$\bbeta^{(t+1)} = \arg \max_{\bbeta} -v^{(t)}_j\frac{1}{2}(\tby - \bX \bbeta)^T(\tby - \bX \bbeta) - \sum^p_{j=1} w_j^{(t)}|\beta_j|$$
where
$$v^{(t)}_j = \frac{a_\delta + (n-1)/2}{b_\delta+1/2(\tby - \bX \bbeta^{(t)})^T(\tby - \bX \bbeta^{(t)})} \text{ and } w_j^{(t)} = \frac{a_j+1}{b_j + |\beta^{(t)}_j|}$$

To test the method, we simulated data using the same as before but letting $\delta^2 \sim IG(a_\delta,b_\delta)$. We then ran 1000 repetitions of the hierarchical adaptive lasso with various settings of $(a_\delta,b_\delta,a,b)$ with the results given in Table \ref{tab:lrdu_hal}. There is clearly more difficulty in estimating the coefficients accurately when the variance of the observations is higher and there is again increased performance with good prior information.

\begin{table}[htp]
\small
\centering
\caption{Results for the HAL (linear regression, random $\delta^2$) \vspace{10pt}}
\label{tab:lrdu_hal}
\begin{threeparttable}
\begin{tabular}{| c | c | c | c | c | c | c |}
\hline
$n$         & $(a_\delta,b_\delta)$ & $(a,b)$       & avg. error      & \% correct   & avg. false positives & avg. false negatives \\
\hline
40          & (3,5) & $(2,0.1)$     & 0.5509          & 90.5         & 0.040          & 0.070     \\
\hline
40          & (1,1) & $(2,0.1)$      & 0.7302          & 79.8         & 0.058          & 0.265   \\
\hline
40          & (1,4) & $(2,0.2)$      & 1.5046          & 53.0         & 0.085          & 0.742   \\
\hline
40          & (1,4) & $(2,0.2)$\tnote{$*$} & 1.1865          & 78.0         & 0.047          & 0.318   \\
\hline
\end{tabular}
\begin{tablenotes}
 \item[$*$] $(a_2,b_2,a_5,b_5) = (2,2,2,2)$
\end{tablenotes}
\end{threeparttable}
\end{table}

\subsubsection{Grouped Variable Selection}
For grouped variable selection, we use $p=32$ with groups of size 4. We let $\beta_{1:4} = (3,1.5,2,0.5)'$, $\beta_{9:12} = (6,3,4,1)'$, $\beta_{17:20} = (1.5,0.75,1,0.25)'$ with all other components set to 0. The groupings of variables were given by $G_i = \{4i + k : k \in \{1,2,3,4\}\}$. As with ungrouped variable selection, the hierarchical adaptive version of the group lasso gives lower average errors and has a higher percentage of correct models chosen compared to the standard group lasso.

\begin{table}[htp]
\small
\centering
\caption{Results for the GLASSO (linear regression, $\delta = 3$) \vspace{10pt}}
\label{tab:lrg_lasso}
\begin{tabular}{| c | c | c | c | c | c |}
\hline
$n$         & $\tau$       & avg. error      & \% correct   & avg. false positives & avg. false negatives \\
\hline
40          & $1/12$       & 3.4738          & 65.4         & 0.580           & 1.012     \\
\hline
40          & $0.1$        & 3.1407          & 70.5         & 0.948          & 0.432   \\
\hline
\end{tabular}
\end{table}

\begin{table}[htp]
\small
\centering
\caption{Results for the GHAL (linear regression, $\delta = 3$) \vspace{10pt}}
\label{tab:lrg_hal}
\begin{tabular}{| c | c | c | c | c | c |}
\hline
$n$         & $(a,b)$       & avg. error      & \% correct   & avg. false positives & avg. false negatives \\
\hline
40          & $(2,0.75)$     & 2.1267          & 89.6         & 0.328          & 0.144     \\
\hline
40          & $(2,0.7)$      & 2.1205          & 91.1         & 0.232          & 0.176   \\
\hline
\end{tabular}
\end{table}

\subsection{Logistic regression}
In logistic regression with $y_i \in \{-1,1\}$, one has $p(\by|\bX,\bbeta) = \prod^n_{i=1}(1+\exp(-y_i\bbeta^T\bx_i))^{-1}$ so the log-likelihood is 
\begin{align}
\label{eqn:ll_logistic}
\log p(\by|\bX,\bbeta) = -\sum^n_{i=1} \log (1+\exp(-y_i\bbeta^T\bx_i))
\end{align}

The Jeffreys prior for this likelihood is given by $p(\bbeta) \propto |X'VX|^{1/2}$, where $V$ is a diagonal matrix with
$$v_{i,i} = \frac{\exp(-\bbeta^T\bx_i)}{[1 + \exp(-\bbeta^T\bx_i)]^2}$$
As a result, the parametrization-invariant MAP estimate requires us to minimize
$$\bbeta_{MAP} = \arg \min_\beta -\log f(\bX|\by,\bbeta) + \frac{1}{2}\log |X'VX| - \log p(\theta)$$
Unfortunately, while $-\frac{1}{2}\log |X'VX|$ is convex, $\frac{1}{2}\log |X'VX|$ is not so the resulting minimization problem is not convex. However, this does not seem to be a a serious issue in our examples as the term $\log |X'VX|$ is relatively constant in the regions of high posterior density and so including the MAP correction has little effect on the results.

To test the method, we simulated data using $\bbeta = (3,1.5,0,0,2,0,0,0)^T$ and $X \sim N(0,\Sigma)$ with $\Sigma_{i,j} = 0.5^{|i-j|}$ as with the linear regression simulations. We then ran 1000 repetitions of the hierarchical adaptive lasso and the standard lasso on this problem with various settings of $(a,b)$ and $\tau$ respectively with the results given in Tables \ref{tab:logr_lasso}-\ref{tab:logr_hal}. An interesting result with this example is that for $(a,b) = (2,0.1)$ except for $(a_2,b_2,a_5,b_5) = (2,2,2,2)$, the HAL gave poor results due to the correlation of the predictors and the relatively high penalization of $\beta_1$. In this case, $\beta_1$ was excluded from the model associated with the MAP estimate in every simulation. Using additionally $(a_1,b_1) = (2,0.5)$ led to a drastic improvement in the results, highlighting the importance the prior hyperparameters can have.

\begin{table}[htp]
\small
\caption{Results for the LASSO (logistic regression) \vspace{10pt}}
\label{tab:logr_lasso}
\centering
\begin{threeparttable}
\begin{tabular}{| c | c | c | c | c | c |}
\hline
$n$         & $\tau$        & avg. error      & \% correct   & avg. false positives & avg. false negatives \\
\hline
80          & 1/7.5           & 2.8559          & 62.1          & 0.387          & 0.098     \\
\hline
80          & 0.1\tnote{*}           & 2.7212         & 93.9          & 0.008           & 0.053   \\
\hline
\end{tabular}
\begin{tablenotes}
 \item[*] $(\tau_2,\tau_5) = (1,1)$
\end{tablenotes}
\end{threeparttable}
\end{table}

\begin{table}[htp]
\small
\caption{Results for the HAL (logistic regression) \vspace{10pt}}
\label{tab:logr_hal}
\centering
\begin{threeparttable}
\begin{tabular}{| c | c | c | c | c | c |}
\hline
$n$         & $(a,b)$       & avg. error      & \% correct   & avg. false positives & avg. false negatives \\
\hline
80          & $(2,0.65)$     & 1.3736          & 65.4         & 0.33          & 0.114     \\
\hline
80          & $(2,0.1)$\tnote{$*$}     & 3.2084          & 0.0         & 0.00         & 1.000   \\
\hline
80          & $(2,0.1)$\tnote{$\dagger$}     & 1.1228          & 99.2         & 0.00         & 0.008   \\
\hline
\end{tabular}
\begin{tablenotes}
 \item[$*$] $(a_2,b_2,a_5,b_5) = (2,2,2,2)$
 \item[$\dagger$] $(a_1,b_1,a_2,b_2,a_5,b_5) = (2,0.5,2,2,2,2)$
\end{tablenotes}
\end{threeparttable}
\end{table}

\subsection{Gaussian graphical models}
The log-likelihood for this model (after standardization) is
$$\log p(X|\Omega) = \frac{n}{2}\log|\Omega| - \frac{n}{2} \text{tr}(S\Omega)$$
where $S = 1/n \sum^n_{i=1}x_i^T \Omega x_i$

Jeffreys prior for this likelihood is given by $p(\Omega) \propto |\Omega|^{(p+1)/2}$ so we can find the parametrization-invariant MAP using
$$\Omega^{(t+1)} = \arg \max_{\Omega} \frac{n-p-1}{2}\log|\Omega| - \frac{n}{2} \text{tr}(S\Omega) - \sum^p_{i=1}\sum^p_{j=i} w_{ij}^{(t)}|\Omega_{ij}|$$
where
$$w_{ij}^{(t)} = \frac{a_{ij}+1}{b_{ij} + |\Omega^{(t)}_{ij}|}$$
In order for the likelihood with the MAP correction term to be concave, we require $n > p + 1$ since $-\log \det$ is a convex function.

To test the method, we simulated data using
$$\mathbf{\Omega} = \left(
\begin{array}
[c]{cccccccc}%
1 & 0 & 0 & 0 & 0.5 & 0 & 0 & 0 \\
0 & 1.5 & 0 & 0.2 & 0.8 & 0 & 0 & 0 \\
0 & 0 & 0.5 & 0.3 & 0 & 0.2 & 0 & 0 \\
0 & 0.2 & 0.3 & 2 & 0 & 0 & 0 & 1.5 \\
0.5 & 0.8 & 0 & 0 & 1 & 0 & 0.5 & 0 \\
0 & 0 & 0.2 & 0 & 0 & 0.5 & 0.3 & 0 \\
0 & 0 & 0 & 0 & 0.1 & 0.3 & 1.5 & 0 \\
0 & 0 & 0 & 1.5 & 0 & 0 & 0 & 2
\end{array}
\right)$$
and again used 1000 repetitions of the procedure using the LASSO and the HAL with the results given in Tables \ref{tab:ggm_lasso}-\ref{tab:ggm_hal}. The HAL clearly has superior performance when using hyperparameters such that the average number of false positives and false negatives are roughly equal.

\begin{table}[htp]
\small
\centering
\caption{Results for the LASSO (GGM) \vspace{10pt}}
\label{tab:ggm_lasso}
\begin{tabular}{| c | c | c | c | c | c |}
\hline
$n$         & $\tau$       & avg. error      & \% correct   & avg. false positives & avg. false negatives \\
\hline
40          & $1/45$       & 4.676          & 23.9         & 2.789           & 1.887     \\
\hline
40          & $1/50$       & 4.22          & 22.3         & 2.081           & 2.139     \\
\hline
\end{tabular}
\end{table}

\begin{table}[htp]
\small
\centering
\caption{Results for the HAL (GGM) \vspace{10pt}}
\label{tab:ggm_hal}
\begin{tabular}{| c | c | c | c | c | c |}
\hline
$n$         & $(a,b)$       & avg. error      & \% correct   & avg. false positives & avg. false negatives \\
\hline
40          & $(1,0.075)$     & 2.594          & 65.4         & 1.304          & 1.290     \\
\hline
40          & $(2,0.1)$     & 2.850          & 57.7         & 1.343          & 1.507     \\
\hline
\end{tabular}
\end{table}

\section{Discussion}
We have proposed a MAP-based variable selection method using a hierarchical prior for $\bbeta$ that works reasonably well in practice and brings together a variety of related approaches in the literature. In particular, the estimate itself corresponds to the solution of a non-convex penalized optimization problem, with properties similar to that in \cite{penlike}, ie. estimates of large coefficients tend to be penalized less than in standard $\ell_1$-penalized optimization approaches, while still being sparse and continuous in the data. A possibly more important contribution is the interpretation the method gives for various methods that have been proposed without Bayesian interpretations, in particular for adaptive or one-step methods in the statistics literature and for iteratively reweighted methods in the machine learning and signal processing literatures. This interpretation allows for manipulation of the hierarchy in application-specific ways. 

A number of open questions remain when using this class of methodology for variable selection. One is how to resolve the issue of multimodality of the posterior due to the non-concavity of the log of the prior density. Another is assessing the utility of point estimates when there is little guarantee that the model corresponding to the MAP estimate has significant posterior mass from a Bayesian variable selection perspective. In this work, we feel these issues are secondary as the major contribution is in the Bayesian interpretation and generalization of increasingly popular penalized optimization methods amongst practitioners.

\section*{Acknowledgments}
We would like to thank Mark Schmidt for the public availability and usability of his $\ell_1$-penalized optimization code, which made the implementation of our methods straightforward.

\bibliographystyle{unsrt}
\bibliography{sip}

\end{document}